\documentclass{emulateapj}
\usepackage{amsmath} 
\usepackage{amssymb} 
\usepackage{verbatim}
\usepackage{graphicx}
\usepackage[percent]{overpic}
\usepackage{hyperref}
\usepackage{subfigure}
\usepackage{esint}
\usepackage{url}
\usepackage{mycommands}
\begin{document}


\title{Spatial Distribution of Pair Production over the Pulsar Polar Cap}



\author{Mikhail A. Belyaev}
\affil{Astronomy Department, University of California, Berkeley, CA 94720 \\ mbelyaev@berkeley.edu; (858) 342-4287}
\and
\author{Kyle Parfrey\altaffilmark{1}}
\affil{Lawrence Berkeley National Laboratory, 1 Cyclotron Road, Berkeley, CA 94720}
\altaffiltext{1}{Einstein Fellow}

\begin{abstract}
Using an analytic, axisymmetric approach that includes general relativity, coupled to a condition for pair production deduced from simulations, we derive general results about the spatial distribution of pair-producing field lines over the pulsar polar cap. In particular, we show that pair production on magnetic field lines operates over only a fraction of the polar cap for an aligned rotator for general magnetic field configurations, assuming the magnetic field varies spatially on a scale that is larger than the size of the polar cap. We compare our result to force-free simulations of a pulsar with a dipole surface field and find excellent agreement. Our work has implications for first-principles simulations of pulsar magnetospheres, and for explaining observations of pulsed radio and high-energy emission. 
\end{abstract}

\section{Introduction}

In their seminal paper, \cite{GoldreichJulian} first demonstrated that the electromagnetic force dominates gravity and particle inertia in the pulsar magnetosphere. Thus, in the presence of a plasma dense enough to screen the component of the electric field parallel to the magnetic field, the electromagnetic force on the magnetospheric plasma approximately vanishes. However, because the plasma is flowing out relativistically along open magnetospheric field lines, it must be constantly replenished. Otherwise a ``vacuum gap", in which the plasma density is too low to screen the parallel electric field, will develop. 

An attractive mechanism for supplying the magnetosphere with plasma is pair production. \cite{Sturrock} proposed that pairs are produced at the polar cap by annihilation of gamma-ray photons on magnetic field lines ($\gamma$--$B$ mechanism). This pair production mechanism has since been extensively studied using both analytical models and simulations \citep{AronsScharlemann,DaughertyHarding,BeskinGR,MuslimovTsygan,HardingMuslimov1998,HibschmanArons,TimokhinArons}. An alternate or potentially complementary mechanism of pair production is annihilation of gamma-ray photons on optical or X-ray photons ($\gamma$--$\gamma$ mechanism), proposed by \citet{ChengHoRuderman}. 

The most compelling observational evidence for pair production comes from observations of synchrotron emission in pulsar wind nebulae. For the Crab, the inferred pair multiplicity based on the injection rate of particles into the nebula needed to explain the synchrotron emission is $\kappa \sim 10^6$ \citep{Shklovsky,deJager,Bucciantini_nebula}. Pair multiplicities in the range $\kappa \sim 10^4$--$10^6$ are also inferred for a host of other pulsar wind nebulae \citep{Arons_review}.

Because nebular observations and theoretical models point to high pair multiplicities, it is often useful to adopt the point of view that pair production is efficient and fills the entire magnetosphere with a dense, highly conductive pair plasma. Additionally, the energy density of this plasma is small compared to the energy density of the electromagnetic field, both inside the light cylinder and in the wind region of open field lines beyond it. 

Under these conditions, the plasma is governed by the equations of degenerate force-free electrodynamics \citep{TM82,MT82,KomissarovFFE}. The force-free approach has been used in simulations to model the topology of magnetic flux surfaces in the magnetosphere and compute the spindown luminosity of the pulsar \citep{Contopoulos,Gruzinov,Timokhin_forcefree,Komissarov,Spitkovsky,McKinney,CK,LiSpitkovsky,Kalapotharakos,Parfreyetal,PetriFF,Tchekhovskoy,PetriGR}.

Although the force-free assumption is a useful simplification, gamma-ray observations inform us that there are regions of the magnetosphere that are not force-free \citep{MAGICa,Fermi2,AGILE,Fermia}. Sites of gamma-ray emission, which are associated with either magnetospheric vacuum gaps or regions of high current density, involve transfer of energy from the electromagnetic field to particles. This is outside the scope of the force-free paradigm, which has no dissipation.

In order to better connect theory with observations, \citet{LiSpitkovsky,Kalapotharakos} pioneered the dissipative force-free approach. \citet{Kalapotharakos_FIDO} showed that assuming a force-free solution inside the light cylinder and a dissipative solution outside provides a better fit to the {\it Fermi} data than assuming a dissipative solution everywhere. 

Global multidimensional, particle in cell (PIC) simulations \citep{PhilippovSpitkovsky,ChenBeloborodov,CeruttiSpitkovsky,Cerutti3D,PhilippovSpitkovsky1,PhilippovGR,BelyaevPIC2,BelyaevPIC} are the natural next step in magnetosphere modeling, because they can self-consistently model plasma instabilities, vacuum gaps, and current sheets. Thus, they provide a handle on the plasma processes leading to dissipation of electromagnetic energy and particle acceleration in the magnetosphere. 

Using 1D PIC simulations of pair production on magnetic field lines at the polar cap, \cite{TimokhinArons} confirmed the \cite{Beloborodov_cloud} result for the conditions under which a polar cap pair cascade is ignited (discussed further in \S \ref{considerations}). Subsequently, these conditions have been confirmed in 2.5D axisymmetric simulations of the aligned rotator with general relativity \citep{PhilippovGR}. 

In this paper, we create an analytical model for the spatial extent of pair-producing field lines over the polar cap using the \cite{Beloborodov_cloud} result. Consistent with \cite{PhilippovGR}, we find that general relativity is a {\it necessary} ingredient for pair production to occur in the aligned rotator over much of the polar cap. The general-relativistic effect enabling polar cap pair production is the dragging of inertial frames \citep{MuslimovTsygan}. 

However, even with general relativity included, pair production occurs over only a fraction of the polar cap. Moreover, whereas the PIC simulations of \cite{PhilippovGR} were specific to a dipole magnetic field, we derive analytical results that apply with much weaker assumptions about the magnetic field at the surface of the star. We mention that \cite{Gralla} have recently published analytical results similar to ours.

The paper is organized as follows. In \S \ref{considerations} we set the framework for our derivation and define conventions. In \S \ref{four_current} we derive a general condition valid for a slowly rotating Kerr metric that allows us to determine the regions of the polar cap over which a pair cascade operates. In \S \ref{jdistribution} we apply our analytical theory to the specific case of a dipole surface magnetic field. In \S \ref{computational} we compare our results with force-free simulations, which show excellent agreement with the analytical theory. In \S\ref{discussion} we discuss the implications of our work for simulations of pulsar magnetospheres and  for understanding the physical origin of pulsed gamma-ray emission from pulsars.

\section{General Considerations}
\label{considerations}

We consider an axisymmetric, force-free pulsar magnetosphere around a neutron star, which has a magnetic moment given by $\bfmu = \mu \zhat$ and rotates with angular frequency $\bfOmega = \Omega \zhat$. By the force-free assumption
\ba
\label{ff_current}
 \rho \bfE + c^{-1} \bfJ \times \bfB = 0,
\ea
and the Lorentz invariants of the electromagnetic field satisfy $\bfE \cdot \bfB = 0$ and $B^2-E^2 > 0$.  

We define the light cylinder radius as $R_{\rm LC} \equiv c/\Omega$. The regions on the pulsar surface containing the footpoints of open magnetospheric field lines are the polar caps, and for typical magnetic field configurations, there are two polar caps. From magnetic flux conservation, the fractional area occupied by both of them combined at the surface of the neutron star is $2 A_{\rm PC}/4 \pi r_*^2 \sim r_*/R_{\rm LC}$, where $A_{\rm PC}$ is the surface area of a single polar cap. For the Crab pulsar, $R_{\rm LC}/r_* \approx 130$, while for millisecond pulsars, $R_{\rm LC}/r_* \sim 10$. In either case, the polar caps occupy a small fraction of the neutron star surface area. For a dipolar field in axisymmetry, the polar caps are both centered on the rotational axis (at opposite poles), in which case the angular extent of a polar cap is $\theta_{\rm PC} \approx \sqrt{A_{\rm PC}/\pi r_*^2} \sim \sqrt{r_*/R_{\rm LC}}$. 

If higher order multipole components are dominant at the surface of the star and generate a complicated small scale field structure (on the scale of the polar cap) then it is difficult to make further conclusions. However, if multipolar effects are on a scale that is much larger than the polar cap size, then progress can be made. In this case, one or both of the polar caps can be significantly displaced from the rotational axis \citep{Arons79}. The angular extent of such a displaced polar cap in axisymmetry can be estimated using magnetic flux conservation as $\theta_{\rm PC} \sim A_{\rm PC}/(2\pi r_*^2 \sin \theta_0) \sim r_*/(R_{\rm LC} \sin \theta_0)$, where $\theta_0 \gtrsim \theta_{\rm PC}$ is the displacement of the center of the annulus from the rotational axis. Notice that for $\theta_0 \gg \theta_{\rm PC}$, the angular extent of the off-axis polar cap is smaller than that of the axis-centered one, even though their surface areas are comparable. Additionally, \citet{HardingMuslimov11} considered non-axisymmetric displacements of the polar cap from the magnetic dipole axis. Such perturbations could be important when considering the 3D non-axisymmetric case.

\subsection{Pair-Production Criteria}

\citet{Beloborodov_cloud} found a simple set of local criteria that determine whether pair production occurs at the polar cap:
\begin{align}
\label{conditions}
J_B/J_{\rm GJ}  < 0 &, \ \ \ \text{pair production} \nn \\
0 < J_B/J_{\rm GJ}  < 1 &, \ \ \ \text{no pair production} \\
J_B/J_{\rm GJ}  > 1 &, \ \ \ \text{pair production} \nn.
\end{align}
Here $J_B$ is the local steady-state current density along a magnetic field line and $J_{\rm GJ}$ is the Goldreich-Julian current density, which is defined as $J_{\rm GJ} \equiv \rho_{\rm GJ} c$, where $\rho_{\rm GJ}$ is the Goldreich-Julian charge density. The sign of $J_B$ is the same as the sign of $\bfJ \cdot \nhat$, where $\nhat$ is the outward-pointing unit vector normal to the surface of the star. It is important to note that $\rho_{\rm GJ}$ is set locally (modulo small rotationally induced changes in the poloidal magnetic field; see Section~\ref{ChargeCurrentDensities}), whereas $J_B$ is determined by the global structure of the magnetosphere, and thus can differ significantly from $J_{\rm GJ}$. Also important is the fact that $J_B$ is the {\it steady-state} i.e. time-averaged current, since the instantaneous current can exhibit rapid variability due to the time-dependent nature of the pair production process.

There is a physical explanation for the pair-production criteria, equation~(\ref{conditions}). When $ 0 < J_B/J_{\rm GJ}  < 1$, the current is carried by the charge density at the surface of the neutron star flowing outward at a velocity less than the speed of light. The resulting flow resembles the steady-state solution predicted by \citet{Mestel_cloud} and \citet{Beloborodov_cloud} and has typical Lorentz factors of order unity (see equation 4 of \citet{Beloborodov_cloud} for maximal Lorentz factor as a function of $J_{\rm GJ}/\rho_{\rm GJ})$. For such low Lorentz factors, pair production on magnetic field lines by either curvature photons or inverse Compton scattered thermal photons from the polar cap is impossible.

Conversely, when $J_B/J_{\rm GJ}  < 0$ or $J_B/J_{\rm GJ}  > 1$, the current {\it cannot} be carried exclusively by the charge density at the surface of the pulsar flowing outward. In this case, a steady-state flow on short timescales is not achieved. Instead, oscillatory bursts of pair production generate plasma pair densities that exceed the Goldreich-Julian value by up to a factor of $\kappa \sim 10^4$--$10^5$ \citep{Timokhin_max}.

\subsection{General Relativity}

For our calculations that involve general relativity, we use a 3+1 split of slowly rotating Kerr spacetime \citep{MT82,membrane}. We work to first order in the dimensionless Kerr spin parameter, $a \equiv J_*c/G M_*^2$, where $J_*$ and $M_*$ are the angular momentum and mass of the neutron star, respectively. The reason for working to first order in $a$ is that frame dragging first appears at this order. Moreover, the problem is simplified because spacetime non-sphericity first enters into the Kerr metric at second order; in other words, in the slow-rotation approximation the frame-dragging rate is a function of the radial coordinate alone. 

The slow-rotation approximation is equivalent to assuming $R_{\rm LC} \gg r_{\rm s}$, where
\ba
r_{\rm s} \equiv \frac{2 GM}{c^2}
\ea
is the Schwarzschild radius. The Kerr metric in this approximation is
\begin{align}
\label{metric}
ds^2 = -\alpha^2 c^2 dt^2 + \alpha^{-2}dr^2 &+ r^2(d\theta^2 + \sin^2 \theta d\phi^2) \\
&-   2\omega_{\rm LT} r^2 \sin^2 \theta d\phi dt \nn,
\end{align}
where the lapse function is defined by $\alpha(r) \equiv \sqrt{1 - r_{\rm s}/r}$. 

The frame dragging rate can be expressed as
\ba
\omega_{\rm LT}(r) &\approx& \frac{2 I_* \Omega G}{c^2 r^3},
\ea
where $I_*$ is the neutron star moment of inertia. Using
\ba
I_* \approx 0.21 \frac{M_* r_*^2}{1 - r_{\rm s}/r_*}
\label{MomentInertia}
\ea
to approximate the neutron star moment of inertia \citep{RavenhallPethick}, we can express the frame-dragging rate at the surface of the star as
\ba
\label{LenseThirring}
\omega_{\rm LT}(r_*) \approx 0.21 \left( \frac{r_{\rm s}/r_*}{1 - r_{\rm s}/r_*} \right) \Omega.
\ea
For a $2 M_\odot$ neutron star with $r_*  = 12$ km, we have $r_{\rm s}/r_* \approx 0.5$ and $\omega_{\rm LT}(r_*)/\Omega \approx 0.21$; for a $1.2 M_\odot$ neutron star with $r_*  = 12$ km, we have $r_{\rm s}/r_* \approx 0.3$ and $\omega_{\rm LT}(r_*)/\Omega \approx 0.09$.

For the the metric in equation (\ref{metric}), Maxwell's equations can be written as
\begin{align}
\label{maxwellkerr}
\bfnabla \cdot \bfE &= 4 \pi \rho \\
\bfnabla \cdot \bfB &= 0 \\
\label{maxwellkerr0}
\bfnabla \times \left(\alpha \bfE + \frac{\bfbeta}{c} \times \bfB \right)  &= -\frac{1}{c} \frac{\partial \bfB}{\partial t}   \\
\label{maxwellkerr1}
\bfnabla \times \left(\alpha \bfB - \frac{\bfbeta}{c} \times \bfE \right)  &= \frac{1}{c} \frac{\partial \bfE}{\partial t} + \frac{4 \pi}{c} \left(\alpha \bfJ - \rho \bfbeta \right).
\end{align}
Physical quantities, such as $\rho$, $\bfJ$, $\bfB$, and $\bfE$ are measured by zero angular momentum observers (ZAMOs) in the frame corotating with absolute space \citep{membrane}. The velocity with which our spherical coordinates rotate relative to this frame is given by the shift vector, $\bfbeta \equiv -\bfomega_{\rm LT} \times \bfr$. The lower-index spatial vector components quoted below are orthonormalized, i.e.\ $A_i \equiv \sqrt{g_{i i}} A^i$, where there is no summation over repeated indices. 

\section{Polar Cap Pair Production: General Results}
\label{four_current}
In this section, we derive general results about the spatial distribution of field lines that support pair production on the polar cap. Our approach is to trace the distribution of current on open field lines in the wind region beyond the light cylinder back to the polar cap, compute the magnitude of the four-current over the polar cap, and apply the pair-production criteria, equation~(\ref{conditions}).  

\subsection{Charge and Current Densities}
\label{ChargeCurrentDensities}
We begin by stating some useful results about charge and current densities for the force-free aligned rotator. The first useful result is that the current flows on magnetic flux surfaces in the magnetosphere. To demonstrate this, consider that in steady state under the assumption of axisymmetry, equation (\ref{maxwellkerr0}) implies that $E_\phi = 0$. As a result, the $\phi$ component of equation (\ref{ff_current}) simplifies to
\ba
\bfJ_{\rm P} \times \bfB_{\rm P} = 0,
\ea
where $\bfB_{\rm P}$ and $\bfJ_{\rm P}$ are the poloidal magnetic field and current, respectively. This means the current flows along magnetic field lines when both are projected into the poloidal plane. 

Moreover, by conservation of magnetic flux,  $\bfnabla \cdot \bfB = 0$, and conservation of charge, $\bfnabla \cdot (\alpha \bfJ - \rho \bfbeta) = 0$, the poloidal current density is proportional to the poloidal magnetic field up to a factor of the lapse function\footnote{Here $\bfnabla \cdot (\rho \bfbeta) = 0$, as azimuthal gradients are zero by axisymmetry and the shift vector is exclusively in that direction.}: $\alpha \bfJ_{\rm P} \propto \bfB_{\rm P}$. We can express this formally by defining a magnetic flux parameter, $\Psi$, which is measured from the pole. In terms of $\Psi$, the poloidal magnetic field is
\ba
\bfB_{\rm P} = \frac{\bfnabla \Psi \times \phat}{2\pi r \sin \theta},
\ea
and the poloidal current density is
\ba
\label{Jp}
\bfJ_{\rm P} = \frac{1}{\alpha}\frac{d I}{d \Psi} \bfB_{\rm P}.
\ea
Here, $I(\Psi)$ is the poloidal current enclosed between the magnetic flux surface $\Psi$ and the pole, as measured using coordinate time, $t$, which coincides with the proper time of static observers at infinity\footnote{Note that this is in general different from the ZAMO's proper time} \citep{membrane}. Note that $(\bfB \cdot \bfnabla) I = 0$, so the enclosed current is constant on flux surfaces. We shall use this fact to map the current in the wind zone back onto the surface of the neutron star when computing the magnitude of the four-current over the polar cap.

The second useful result is a calculation of the charge density in the magnetosphere. To begin, we define a ``corotational velocity"
\begin{align}
\label{V0eq}
\bfV_0 \equiv \begin{cases} 
      \bfOmega \times \bfr, & \text{flat}\\
      \alpha^{-1} \left(\bfOmega - \bfomega_{\rm LT} \right) \times \bfr, & \text{Kerr}
      \end{cases},
\end{align}
in terms of which the electric field is given by 
\ba
\label{corotE}
\bfE = -\bfV_0 \times \bfB/c. 
\ea
This expression is Ferraro's isorotation law, which was generalized to a Kerr spacetime by \cite{BlandfordZnajek}. The value of $\Omega$ on a given magnetic flux surface is set at the footpoint of the flux surface on the neutron star. Thus, the angular frequency is in general a function of the magnetic flux parameter [i.e. $\Omega(\Psi)$], but for simplicity we have assumed that the neutron star rotates as a solid body (i.e. $\Omega$ is a constant).

It is worth pointing out that the corotational velocity {\it is not a physical velocity} associated with the magnetospheric plasma, and thus is not required to be less than the speed of light. Rather it is a pattern velocity that can be interpreted as the ``velocity of a magnetic field line", so one can view the magnetic field as corotating with the pulsar. The actual drift velocity of the plasma, 
\ba
\label{vdrift}
\bfV_D &=& c \bfE \times \bfB/B^2,
\ea
remains less than the speed of light, since $B^2-E^2 > 0$.

Having defined the electric field in terms of the corotational velocity, the Goldreich-Julian charge density is
\begin{align}
\rho_{\rm GJ} &\equiv - \bfnabla \cdot \left(\bfV_0 \times \bfB \right)/4\pi c \\
&= \frac{-\bfB \cdot \left(\bfnabla \times \bfV_0 \right) + \bfV_0 \cdot \left(\bfnabla \times \bfB \right)}{4\pi c} \\
\label{rhoGJ}
&= - \frac{(\bfOmega - \bfomega_{\rm LT}) \cdot \bfB}{2 \pi c \alpha} + \frac{\bfV_0 \cdot \left(\bfnabla \times \bfB \right)}{4\pi c}.
\end{align}
This expression for the charge density is the same as in \cite{MuslimovTsygan}.

At the polar cap, the first term on the right hand side of equation (\ref{rhoGJ}) is dominant compared to the second as long as the characteristic length scale over which the poloidal magnetic field varies is larger than the polar cap size. For a dipole field, this condition is easily satisfied, since the radius of curvature at the polar cap is $r_C \sim \sqrt{r_* R_{\rm LC}}$, whereas the size of the polar cap is $d_{\rm PC} \sim r_* \sqrt{r_*/R_{\rm LC}}$, so $r_C/d_{\rm PC} \sim R_{\rm LC}/r_* \gg 1$. From now on, we assume that the characteristic length scale of the poloidal magnetic field at the polar cap is larger than the polar cap size, so we can approximate the Goldreich-Julian density using only the first term in equation (\ref{rhoGJ}).

Having derived an expression for the charge density, the next step is to calculate the current density over the polar cap. Because the poloidal current flows on magnetic flux surfaces we can remap the poloidal current from the wind zone beyond the light cylinder onto the polar cap. In particular, we have at the polar cap
\begin{align}
 J^{\mu}J_{\mu} &= -(\rho_{\rm GJ} c)^2 + J_P^2 + J_\phi^2 \\
 &\approx -\left(\frac{(\bfOmega - \bfomega_{\rm LT})\cdot \bfB }{2 \pi \alpha}\right)^2 + J_P^2 + J_\phi^2. \\
  \label{Jmueq}
 &\approx -\left(\frac{(\bfOmega - \bfomega_{\rm LT})\cdot \bfB }{2 \pi \alpha}\right)^2 + \left(\frac{1}{\alpha} \frac{dI}{d\Psi} B_{\rm P}\right)^2.
\end{align}
The first approximation involves ignoring the second term for the charge density in equation (\ref{rhoGJ}). The second approximation is to set $J_\phi \approx 0$, which will not strongly affect the four-current magnitude as long as $|J_\phi/\rho_{\rm GJ}| \sim \Omega r_* \sin \theta \ll c$ on the polar cap. The only thing left to do in order to determine the four-current magnitude over the polar cap is to specify $I(\Psi)$. Since $I(\Psi)$ is invariant on flux surfaces, this is most easily done in the wind zone.  

The poloidal magnetic field lines in the wind zone of the pulsar magnetosphere open up and become radial. Recently, \citet{Tchekhovskoy_current} have found an analytic fit to the distribution of poloidal field lines from force-free simulations of a dipole magnetosphere. Keeping the two largest terms in their approximation, the poloidal field in the wind zone is given by
\begin{align}
\label{poloidalB}
\bfB_{\rm P} \approx \frac{\mu k}{R_{\rm LC}r^2}(1 + A_1(\cos \theta-1)) \rhat,
\end{align}
where $\mu$ is the dipole moment of the neutron star, $k$ is a constant of order unity, and $A_1 \approx 0.22$. Near the polar axis, the poloidal field is well-described by a split monopole structure \citep{Michelmono}, which has $\partial B_r/\partial \theta = 0$. However, it deviates from a split monopole near the equator due to the presence of a distributed return current. This correction is captured in equation (\ref{poloidalB}) via the $A_1(\cos\theta - 1)$ term.

\subsection{Split-Monopole Wind Zone}
\label{splitmonowind}
We begin by considering a split-monopole distribution of poloidal field lines in the wind zone ($A_1 = 0$ in equation (\ref{poloidalB})). The split monopole provides an accurate description near the polar axis and the resulting equations are substantially simpler and more intuitive. However, it does not contain the distributed return current, and we shall keep this in mind during our discussion. We will consider the general case of $\partial B_r/\partial \theta \ne 0$, which captures the distributed return current, in \S \ref{windzonegen}.

The split monopole in flat spacetime has a simple analytical description for the force-free electromagnetic fields and  currents. In the upper half plane ($z > 0$) it is described by the following expressions for the electromagnetic fields and current density:
\begin{align}
\label{monosteady}
B_r &= B_0 \left(\frac{r_0}{r}\right)^2,  &E_r &= 0, &J_r &= -\frac{\Omega}{2 \pi} B_r  \cos \theta \\
B_\theta &= 0,  &E_\theta &=B_\phi, &J_\theta &= 0 \nn \\
B_\phi &= -B_r \frac{r \sin \theta}{R_{\rm LC}} , &E_\phi &= 0, &J_\phi &= 0. \nn
\end{align}
The solution in the lower half plane, $z < 0$, is the same but replacing $B_r \rightarrow -B_r$ so all non-zero values in equation (\ref{monosteady}) change sign. The lower and upper half planes are separated by a current sheet, which carries the return current necessary to prevent charging of the neutron star in steady state.  We also mention that \citet{Lyutikov} studied a split-monopole field in a slowly rotating Kerr spacetime. However, the flat spacetime solution will suffice for our calculations, since the light cylinder is at a sufficiently large radius in all observed pulsars that general relativity is not important beyond it.  

In the split-monopole region, we can use equations (\ref{monosteady}) together with equation (\ref{Jp}) and $\alpha \approx 1$ to write 
\ba
\label{current_split_mono}
\frac{dI}{d\Psi} = -\frac{\Omega}{2\pi} \hat{b}_z  \ \ \ \text{(split monopole)},
\ea
where $\hat{b}_z \equiv \bhat_{\rm P} \cdot \zhat$ and $\bhat_{\rm P}$ is the unit vector along the direction of the poloidal magnetic field. Thus, $\hat{b}_z$ measures the orientation of the poloidal magnetic field with respect to the vertical direction; the reason for this use of notation shall become apparent shortly.

Imagine now tracing a field line from the far-field split-monopole region back to the polar cap. Because $dI/d\Psi$ is constant on magnetic flux surfaces, we can use equation (\ref{current_split_mono}) to write the four-current magnitude given by equation (\ref{Jmueq}) as
\begin{align}
\label{kerr_estimate}
&J^{\mu}J_{\mu} = \left(\frac{B_{\rm P} \Omega}{2\pi \alpha}\right)^2 \left[ \left. \left.  - \left(1 - \frac{\omega_{\rm LT}}{\Omega}\right)^2 \hat{b}^2_z  \right |_{PC} + \hat{b}_z^2 \right |_{SM} \right],
\end{align}
where vertical bars with subscripts ``PC" and ``SM" denote evaluation of a quantity {\it on the same magnetic flux surface} at a point PC on the polar cap and at a point SM in the split-monopole wind zone. The prefactor in front of the square brackets is evaluated at the polar cap.

We now use equation (\ref{kerr_estimate}) to make general statements about spacelike versus timelike four-current regions at the inner edge of the polar cap. This is the edge of the polar cap nearest to the polar axis in $\theta$, where the split-monopole solution is valid. We point out that the polar cap itself can still be displaced away from the polar axis. 

For the split monopole, $J_{\rm GJ}$ is in the same sense as $\rho_{\rm GJ}$ flowing away from the pulsar. In this case, according to the criteria (\ref{conditions}), whether there is pair production on a field line near the inner edge can be reformulated in terms of the four-current magnitude as 
\begin{align}
\label{conditions2}
J^\mu J_\mu  > 0 &, \ \ \ \text{pair production} \nn \\
J^\mu J_\mu  < 0 &, \ \ \ \text{no pair production}.
\end{align}
This expression is valid for both a flat spacetime and a Kerr spacetime given that $\rho$ and $\bf J$ are measured by ZAMOs \citep{KomissarovGR}. 
\subsubsection{Flat Spacetime, Axis-Centered Polar Cap}

We begin by considering the simplest case of a polar cap that is centered on the rotational axis in flat spacetime. In this case, whether the four-current is spacelike or timelike at a point on the polar cap can be determined by comparing the angle the poloidal magnetic field makes with the polar axis at the polar cap and in the split-monopole wind region: 
\begin{align}
  J^{\mu}J_{\mu} \ \text{is} \ \begin{cases} 
      \text{timelike}, & \left. \hat{b}_z^2 \right |_{PC} > \left. \hat{b}_z^2 \right |_{SM}\\
      \text{null}, & \left. \hat{b}_z^2 \right |_{PC} = \left. \hat{b}_z^2 \right |_{SM}\\
      \text{spacelike}, & \left. \hat{b}_z^2 \right |_{PC} < \left. \hat{b}_z^2 \right |_{SM}
   \end{cases}.
   \label{flat4}
\end{align}
If the angle the poloidal magnetic field makes with the vertical ($\zhat$ direction) is less/greater at the polar cap than in the wind zone on the same magnetic flux surface, then the four-current magnitude at the polar cap will be timelike/spacelike. This gives an intuitive way of determining whether the current will be spacelike or timelike in the split-monopole approximation and is a useful feature of the model.

From these considerations, it immediately follows that the four-current on the polar axis is null. This is because for an axis-centered polar cap, the poloidal magnetic field on the polar axis points in the $\zhat$ direction everywhere. Thus, $\hat{b}_z^2 = 1$ in both the wind zone and on the polar cap. Additionally, the four-current magnitude away from the polar axis on the polar cap is timelike in flat spacetime. This is because flux surfaces generally bend away from the polar axis if the polar cap is centered on it, so that $\hat{b}^2_z|_{PC} > \hat{b}^2_z|_{SM}$. These considerations are {\it valid in general} at the inner edge of the polar cap, since the split-monopole solution provides a good approximation in this region.  

\subsubsection{Curved Spacetime, Axis-Centered Polar Cap}
\label{kerr}

The generalization of the criteria (\ref{flat4}) to a slowly rotating Kerr spacetime is given by
\begin{align}
  J^{\mu}J_{\mu} \ \text{is} \ \begin{cases} 
      \text{timelike}, & \left. \left[\left(1 - \frac{\omega_{\rm LT}}{\Omega}\right) \hat{b}_z \right]^2 \right |_{PC} > \left. \hat{b}_z^2 \right |_{SM}\\
      \text{null}, & \left. \left[\left(1 - \frac{\omega_{\rm LT}}{\Omega}\right) \hat{b}_z \right]^2 \right |_{PC} = \left. \hat{b}_z^2 \right |_{SM}\\
      \text{spacelike}, & \left. \left[\left(1 - \frac{\omega_{\rm LT}}{\Omega}\right) \hat{b}_z \right]^2 \right |_{PC} < \left. \hat{b}_z^2 \right |_{SM}
   \end{cases}.
   \label{kerr4}
\end{align}

Comparing equations (\ref{flat4}) and (\ref{kerr4}), we see that the effect of frame dragging is to effectively reduce the magnitude of the charge density at the polar cap as compared to the poloidal current density. This means the four-current near the polar axis is no longer null, but is {\it spacelike}. However, because field lines bend away from the polar axis, it is a local maximum, and the four-current magnitude decreases as we go away from the polar axis.

Additionally, on field lines that have zero poloidal current (for the split-monopole this means $\hat{b}_z|_{SM} = 0$) the four-current magnitude will generally be timelike. In the split-monopole solution the field line of zero poloidal current is the last open field line. However, due to the presence of the distributed return current, the field line of zero poloidal current is shifted inward in the numerical force-free solutions so that it lies between the inner and outer edges of the polar cap. Nevertheless, the fact that the solution contains a field line of zero poloidal current and the fact that the poloidal current varies smoothly across the polar cap means there is a region of timelike four-current at mid-latitudes on the polar cap with $J_{\rm GJ}$ in the same sense as $\rho_{\rm GJ}$ flowing outward. {\it This region does not support pair production}.

\subsubsection{Off-Axis Polar Cap}
If higher order multipoles are important, then the polar cap can be significantly displaced away from the rotational axis \citep{Arons79,Gralla}. Assuming the field varies on a scale that is large compared to the scale of the polar cap, and the dipole component dominates at the light cylinder, we can use the general formalism we have developed. 

The direction of the magnetic field in the off-axis case (and hence $\hat{b}_z$ at the polar cap) depends on the relative strengths of the multipole components. However, from equation (\ref{flat4}) it is clear there is generally a region of spacelike four-current at the inner edge of the polar cap (the edge closer to the rotational axis), even in flat spacetime. This is true as long as the polar cap magnetic field is not exactly parallel to the vertical direction, $(\hat{b}_z |_{PC})^2 \ne 1$. This differs from the axis-centered case for which the field on the axis points vertically by axisymmetry, resulting in a null four-current density. 

General-relativistic frame dragging reduces the Goldreich-Julian density over the polar cap and enlarges the region of spacelike four-current. However, even in the off-axis case with general relativity, there is generically a region of timelike four-current. The reason, as before, is that both the model and the numerical solution (\S \ref{computational}) contain field lines on which the poloidal current goes to zero. Only in the unlikely special case when the magnetic field is at a right angle to the rotational axis ($\hat{b}_z|_{PC} = 0$), implying $\rho_{\rm GJ} = 0$, will the four-current magnitude over the entire polar cap be spacelike, except on the field line that carries no current, where it will be null.

\subsection{Generalized Wind Zone Solution}
\label{windzonegen}

In this section, we relax the split monopole approximation and derive $dI/d\Psi$ for more general distributions of the poloidal flux. Before we can derive $dI/d\Psi$, however, we must make one additional approximation which is that $E_\theta = B_\phi$ in the wind zone beyond the light cylinder. This means the drift velocity of the plasma is radial ($B_\phi \gg B_r$) and equals the speed of light. With this approximation, it is possible to show that the current in the wind zone is null, $J^\mu J_\mu = 0$, and is given by 
\begin{align}
\bfJ_{\rm P} &= \rho_{\rm GJ} c \rhat \\
&=  \left(-\frac{\bfOmega \cdot \bfB}{2 \pi} + \frac{\bfV_0 \cdot \left(\bfnabla \times \bfB \right)}{4\pi} \right) \rhat \\
\label{jpgen}
&= -\frac{\Omega B_r}{2 \pi}\left(\hat{b}_z + \frac{\sin \theta}{2 B_r} \frac{\partial B_r}{\partial \theta} \right) \rhat.
\end{align}
In going from the first line to the second, we have used equation (\ref{rhoGJ}), and in going from the second line to the third we have used the fact that poloidal field lines are radial in the wind zone. This implies $\hat{b}_z = \cos \theta$, but we have written the poloidal current in the form above to make a more direct connection to the discussion in \S \ref{splitmonowind}. 

Combining equations (\ref{Jp}) and (\ref{jpgen}), and using the fact that $\alpha \approx 1$ in the wind zone, we can write
\ba
\label{ipsigen}
\frac{d I}{d \Psi} = -\frac{\Omega}{2 \pi}\left(\hat{b}_z + \frac{\sin \theta}{2 B_r} \frac{\partial B_r}{\partial \theta} \right).
\ea
Equation (\ref{ipsigen}) is the generalization of equation (\ref{current_split_mono}), which is only valid for the special case of the split monopole ($\partial B_r/\partial \theta = 0$). Substituting the expression for $dI/d\Psi$ from equation (\ref{ipsigen}) into equation (\ref{Jmueq}), the four-current magnitude over the polar cap is given by
\begin{multline}
\label{kerr_gen}
J^\mu J_\mu = \left(\frac{\Omega B_{\rm P}}{2 \pi \alpha}\right)^2\left[ \left. -\left(1 - \frac{\omega_{\rm LT}}{\Omega} \right)^2\hat{b}^2_z  \right |_{PC} \right. \\
 \left. \left. + \left(\hat{b}_z + \frac{\sin \theta}{2 B_r} \frac{\partial B_r}{\partial \theta} \right)^2 \right |_{WZ} \right].
\end{multline}
Here ``WZ" stands for wind zone analogous to ``SM" for split-monopole in \S \ref{splitmonowind}. As before, vertical bars denote evaluation on the same magnetic flux surface at the polar cap and in the wind zone, respectively, and the prefactor in front of the square brackets is evaluated at the polar cap.

Comparing equation (\ref{kerr_gen}) with equation (\ref{kerr_estimate}), we see that deviations from the split-monopole, such as the distributed return current, are contained entirely within the term proportional to $\partial B_r/\partial \theta$. Additionally, just like in the case of the split monopole, the lapse factor $\alpha$ simply rescales the four-current magnitude, meaning it does not have an effect on the spatial distribution of pair production. This is in contrast to frame dragging, which makes the four-current magnitude more spacelike by reducing the magnitude of the charge density term relative to the current density term.

We can also understand the reason why the split-monopole solution works well near the inner edge of the polar cap. The term proportional to $\partial B_r/\partial \theta$ which is not present in the split-monopole solution has a prefactor of $\sin \theta$ and thus is suppressed near the inner edge of the polar cap. This suppression is further enhanced by the fact that $\partial B_r/\partial \theta \propto \sin \theta$ for the leading correction to the split-monopole solution (equation \ref{poloidalB}). 

On the other hand, at the outer edge of the polar cap near the last open field line, $ \hat{b}_z |_{WZ} \sim 0$. Thus, the $\partial B_r/\partial \theta$ term dominates here, and the split-monopole solution is no longer a good approximation.

\section{Pair Production: Dipole Surface Field}

In the previous section, we derived general results that allowed us to identify regions of the polar cap supporting pair production. In this section, we consider a specific magnetic field structure at the polar cap and in the wind zone. This will allow us to check our analytical results against simulations in \S \ref{computational}.

For the poloidal magnetic field in the wind zone, we will concurrently consider the split-monopole model and the more accurate distribution described by equation (\ref{poloidalB}). Although the magnetic field at the polar cap is not known in general, modeling of pulsar radio profiles suggests that it may be well-described by a dipole down to the surface of the neutron star \citep{Rankin}. Thus, we assume a dipole surface magnetic field in this section, in line with most pulsar studies.  

\citet{WassermanShapiro} derive the form of a potential magnetic field in the Schwarzschild metric that is dipole for $r \gg r_*$. Defining the dimensionless variable $x \equiv r/r_{\rm s}$, the nonzero magnetic field components are given by
\begin{align}
\label{GR_dipole}
B_r &= \frac{2 \mu \cos \theta }{r^3} f(x), \ \ \ B_\theta = \frac{\mu \sin \theta }{r^3}g(x), \\
f(x) &\equiv -3x^3 \ln(1-x^{-1}) - 3x^2(1+x^{-1}/2), \nn \\
g(x) &\equiv 6x^3 (1-x^{-1})^{1/2} \ln(1-x^{-1}) + \frac{6 x^2 (1-x^{-1}/2)}{(1-x^{-1})^{1/2}} \nn \,.
\end{align}
For $x \gg 1$, the functions $g(x) \rightarrow 1$ and $f(x) \rightarrow 1$, so we recover the field of a dipole with magnetic moment $\mu$.

We can use equation (\ref{kerr_estimate}) to compute the magnitude of the four-current over the polar cap. However, we first need to evaluate the terms in equation (\ref{kerr_gen}) at the polar cap and in the wind zone. For a polar cap field given by equation (\ref{GR_dipole}), $\hat{b}_z$ as a function of $\theta$ is given by
\begin{align}
\label{zhat_bhat}
\left. \hat{b}_z \right|_{PC} = \frac{2 f_* \cos^2 \theta  - g_* \sin^2 \theta }{\sqrt{\left(2 f_* \cos \theta \right)^2 + \left(g_* \sin \theta \right)^2}},
\end{align}
where $f_* \equiv f(x_*)$ and $g_* \equiv g(x_*)$. Using equation (\ref{poloidalB}), we have for the wind zone\begin{align}
\label{zhat_bhat2}
\left. \hat{b}_z \right|_{WZ} &= \cos \theta \\
\label{extra_term}
\left. \frac{\sin \theta}{2 B_r}  \frac{\partial B_r}{\partial \theta}\right|_{WZ} &=
      \frac{A_1 \sin^2\theta}{2(1+A1(\cos\theta -1))}.
\end{align}
For a split-monopole wind zone, the term in equation (\ref{extra_term}) vanishes since $A_1 = 0$ in this case.

The quantities in equations (\ref{zhat_bhat})-(\ref{extra_term}) must be evaluated on the same magnetic flux surface (i.e.\ at the same value of $\Psi$) at the polar cap and in the wind zone. Thus, we need an expression for $\theta(\Psi)$ [or, equivalently, $\Psi(\theta)$] at both the polar cap and in the wind zone. To compute $\Psi(\theta)$ for the split monopole, we first need to determine the normalization of $B_r$ in the wind zone. 

The wind zone magnetic field is given by equation (\ref{poloidalB}) up to a constant of order unity, $k$. This constant specifies the total amount of open flux, or equivalently, the spin down luminosity. For a given value of $k$, the amount of open flux and the spindown luminosity will depend (weakly) on the value of $A_1$. It will turn out, however, that our results are insensitive to the exact value of $k$ as long as it is of order unity. Thus, we leave it as a free parameter in the model that we can vary. 

To gain some intuition for why $k$ is of order unity it is useful to consider a split-monopole magnetic field ($A_1 = 0$), and we shall refer to $k$ as $k_{\rm SM}$ when explicitly referencing the split-monopole distribution of field lines. Setting $k_{\rm SM} = 1$ makes the total magnetic flux in the split monopole wind-zone region equal to the total magnetic flux beyond the light cylinder for a surface dipole field in flat spacetime. However, a value of $k_{\rm SM} \approx 1.5^{1/2}$ is necessary for the spindown luminosity to agree with force-free simulations using a surface dipole field in flat spacetime. We measure from simulations values of $k_{\rm SM}$ for various light cylinder radii in flat and curved spacetimes in \S \ref{computational} and find them always to be near unity ($1 < k_{\rm SM} < 2$). 

We can express $\Psi(\theta)$ in the wind zone and at the polar cap as
\begin{align}
\label{psi_theta}
\left. \Psi(\theta) \right|_{WZ}  &=  \frac{2 \pi \mu k}{R_{\rm LC}}\left[1-\cos \theta -\frac{A_1}{2}(1 - \cos\theta)^2 \right]\\
\left. \Psi(\theta) \right|_{PC} &=  \frac{2 \pi \mu f_*}{r_*} \sin^2 \theta.
\end{align}
Defining the dimensionless variables
\begin{align}
\psi \equiv \frac{R_{\rm LC}}{ 2\pi \mu k} \Psi, \ \ \ \overline{\psi} \equiv \frac{r_*}{2 \pi \mu f_*} \Psi,
\end{align}
we can solve for $\cos \theta$ in the wind zone and $\sin^2 \theta$ at the polar cap in terms of $\psi$ and $\overline{\psi}$, respectively:
\begin{align}
\label{wz_theta}
\cos \theta &= 1-\frac{1-\sqrt{1-2A_1\psi}}{A_1} & (\text{wind zone})\\
\label{pc_theta}
\sin^2\theta &= \overline{\psi} & (\text{polar cap}).
\end{align}
Note that the polar cap on the surface of the neutron star is defined by values of $\psi$ in the range $0 \le \psi < 1$. Also, for a split-monopole wind zone we can take $A_1 \rightarrow 0$ to derive the simpler expression $\cos \theta = 1 - \psi$.

From here, the procedure to determine the four-current magnitude over the polar cap is straightforward, albeit mechanically involved, so we simply outline it. By substituting expression (\ref{wz_theta}) into equations (\ref{zhat_bhat2}) and (\ref{extra_term}) and expression (\ref{pc_theta}) into equation (\ref{zhat_bhat}), one can derive a formula for the four-current magnitude as a function of $\Psi$. By construction, the evaluation of the terms in equation (\ref{kerr_gen}) can then be performed on the same magnetic flux surface, which allows one to derive the four-current magnitude as a function of $\Psi$. One can then use equation (\ref{pc_theta}) to express the four-current magnitude as a function of $\theta$ over the polar cap.

\begin{figure}
\centering
\includegraphics[width=.49\textwidth]{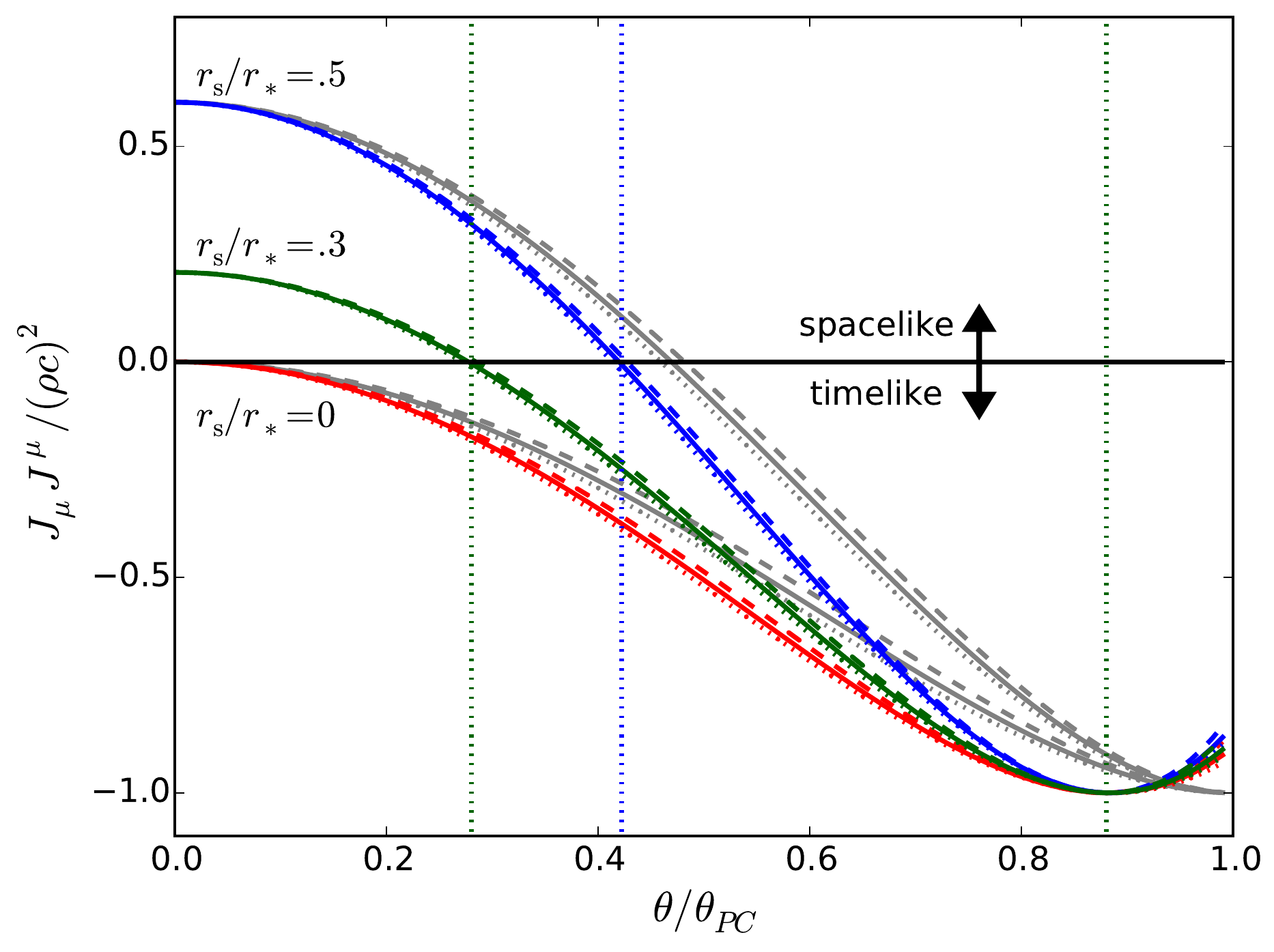}
\caption{Analytically computed four-current magnitude normalized by $(\rho c)^2$ as a function of angle measured from the polar axis for a dipole surface field. The red set of curves is for a flat spacetime and the green and blue sets of curves are for slowly rotating Kerr spacetimes with $r_{\rm s}/r_* = 0.3$ and $r_{\rm s}/r_* = 0.5$, respectively. All of the color curves use equation (\ref{poloidalB}) to approximate the poloidal magnetic field in the wind zone; the upper/lower sets of gray curves are analogous to the blue/red curves, but assuming a split-monopole distribution for the poloidal magnetic field in the wind zone. The split-monopole is a good approximation near the pole, but is inaccurate near the last open field line. Within each bundle of curves, the dotted line is for the parameters $(R_{\rm LC}/r_* = 100,k_{\rm SM} = 1)$, the solid line is for the parameters $(R_{\rm LC}/r_* = 10,k_{\rm SM} = 1)$, and the dashed line is for the parameters $(R_{\rm LC}/r_* = 10,k_{\rm SM} = 2)$.}
\label{flat_fig}
\end{figure}

Fig.~\ref{flat_fig} shows $J^{\mu}J_{\mu}/(\rho c)^2$ as a function of $\theta/\theta_{\rm PC}$, where $\theta_{\rm PC}$ is the angular extent of the polar cap. In flat spacetime, the four-current is timelike over the entire polar cap, except on the polar axis where it is null. On the other hand, in a slowly rotating Kerr spacetime there is a spacelike four-current region that includes the polar axis, as evidenced by the blue and green curves. For each of the blue and green curves, we have denoted the pair production boundaries using vertical dotted green and blue lines, respectively. Note that the rightmost blue and green dotted lines are on top of each other, so they appear as a single line. 

We can split the polar cap into different regions, which have the dotted lines as their boundaries. In region 1, which extends to the right from the polar axis to the first dotted line, there is pair production, because the four-current is spacelike. In region 2, between the two dotted lines, there is no pair production because the four-current is timelike and $0 < J_B/\rho_{\rm GJ}c < 1$. Region 3, which extends to the right from the second dotted line to the last open field line, is the region of distributed return current. Although the four-current is timelike in this region, we have $J_B/\rho_{\rm GJ}c < 0$, so there is pair production. 

Note also that the compactness of the neutron star has a drastic effect on the location of the boundary between regions 1 and 2. In particular as $r_{\rm s}/r_* \rightarrow 0$, the size of region 1 shrinks and region 1 disappears entirely in flat spacetime. This is because the region of pair production near the polar axis is due entirely to the frame-dragging effect, which is absent in flat spacetime. On the other hand, we see that varying the compactness has virtually no effect on the location of the boundary between regions 2 and 3. Thus, the size of the distributed return current region scaled to the polar cap size is not affected by general relativity. 

We also see from Fig. \ref{flat_fig} that varying the dimensionless parameters $k_{\rm SM}$ and $R_{\rm LC}/r_*$ has very little effect on the distribution of the four-current magnitude over the polar cap. However, it does affect the size of the polar cap, since the polar cap becomes smaller as the compactness of the neutron star increases \citep{Gralla} or as the light cylinder moves further out. However, by normalizing $\theta$ by $\theta_{\rm PC}$, we have removed the size-variation of the polar cap so we can compare the distribution of four-current magnitude across the polar cap for different parameter regimes directly. 

\label{jdistribution}

\newpage 

\section{Computational Results}
\label{computational}
The analytical models have been formulated in the slow-rotation approximation to the Kerr metric, equation~(\ref{metric}).  We have performed a set of time-dependent axisymmetric simulations of the pulsar magnetosphere in the complete Kerr spacetime in order to verify the models at low spins, and investigate their applicability to rapidly rotating pulsars.  While the Kerr solution does not describe precisely the spacetime outside an extended rotating body, we expect it to be sufficiently realistic for our present purposes. 


In the simulations, the equations of general-relativistic force-free electrodynamics are solved by the \textsc{phaedra} code \citep{Parfreyetal, ParfreyPhDT}, using the Kerr-Schild spacetime foliation and spherical coordinates.  The compactness parameter is set to $r_{\rm s}/r_* = 0.5$ for all runs. The computational domain consists of the space $r_* \leq r \leq 2,\!000 r_*$ and $0 < \theta < \pi$. The grid has $N_r \times N_\theta = 1024 \times 512$ nodes, which are concentrated near the star by a smooth coordinate mapping.

For each simulation, we set the light cylinder radius $R_{\rm LC}$. The corresponding dimensionless spin parameter is given by $a = J_*c/G M_*^2$, where the star's angular momentum is $J_* = I_* \Omega_*$; equations~(\ref{LenseThirring}) and (\ref{MomentInertia}) give
\ba
a = 0.21 \frac{r_*^2}{1 - r_{\rm s}/r_*}  \frac{1}{R_{\rm LC} \, r_{\rm g}},
\ea
where $r_{\rm g}$ is the star's gravitational radius, $r_{\rm g} = G M_*/c^2 = r_{\rm s} / 2$.

We performed five simulations: two in the Kerr metric with $R_{\rm LC}/r_* = 5$ and 10, giving $a = 0.336$ and 0.168 respectively; two in flat spacetime with the same $R_{\rm LC}/r_*$, and one in the Schwarzschild metric ($a = 0$) with $R_{\rm LC}/r_* = 5$. The Wasserman-Shapiro dipole, equation~(\ref{GR_dipole}), was used as the initial field configuration for the Schwarzschild simulation; in the Kerr metric this field must be adjusted to respect solenoidality [see e.g.\ \citet{ParfreyPhDT}].

\begin{figure}
\centering
\includegraphics[width=3.4in]{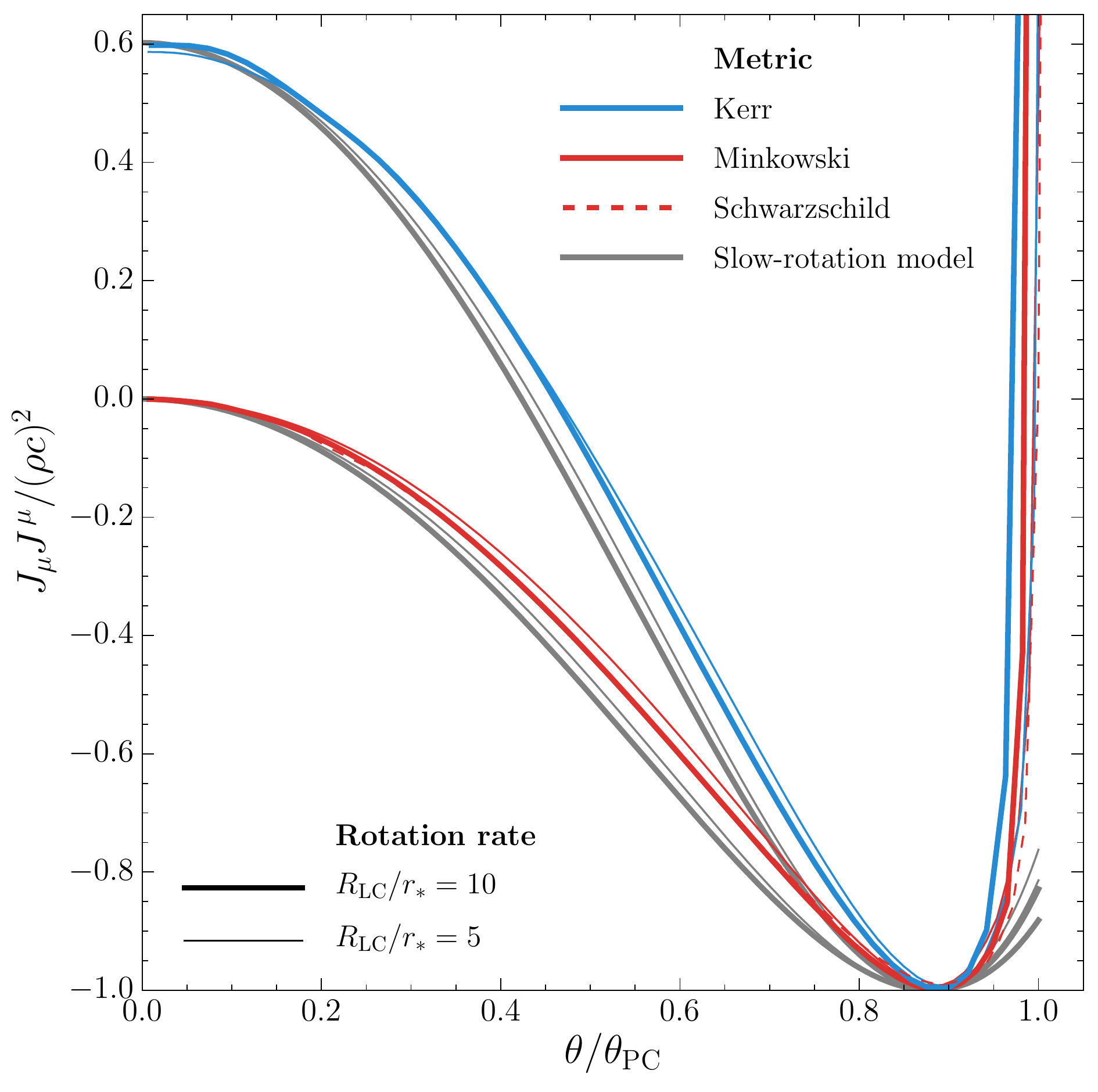}
\caption{Simulation results: normalized four-current magnitude across the polar cap, in the Kerr (solid blue, $r_{\rm s}/r_* = 0.5$), Minkowski (solid red), and Schwarzschild (dashed red, $r_{\rm s}/r_* = 0.5$) spacetimes. The gray curves show the analytical model corresponding to each simulation, using a two-term approximation for the distribution of the poloidal magnetic field in the wind zone. Thick (thin) lines are drawn for $R_{\rm LC}/r_* = 10$ ($R_{\rm LC}/r_* = 5$).  }
\label{sim_fig}
\end{figure}

The normalized four-current magnitude, $J_\mu J^\mu/\rho c^2$, over the polar cap is shown in Fig.~\ref{sim_fig}; note that, as above, the charge density $\rho$ is defined in the rest frame of the observer with zero radial velocity and corotating with spacetime at the Lense-Thirring angular velocity. In this figure, results from the simulations are shown in blue and red, and the model curves are drawn in gray; the model corresponds to the two-term poloidal flux distribution of equation~(\ref{poloidalB}).

For the simulation curves, the normalizing polar cap extent $\theta_{\rm PC}$ is determined by equating the magnetic flux through the polar cap to the open flux in the wind zone $\Psi_{\rm open}$, which is in turn defined as the total flux through the hemisphere $0 < \theta < \pi/2$ at $r = 2 R_{\rm LC}$. The open flux can be directly found from the simulation once the magnetosphere has reached a steady state. 

The analytic models require a value of $k$, which sets the open flux through the polar cap. We fix this value for each model such that the open flux in the model and the corresponding simulation are the same. Specifically, we use the open flux $\Psi_{\rm open}$ from the corresponding simulation and equation~(\ref{psi_theta}) with $A_1 = 0.22$: $k = \Psi_{\rm open} R_{\rm LC} / \left[2 \pi \mu \left(1 - A_1/2 \right)\right]$. For comparison, we also report results for the simpler split-monopole model, for which $k_{\rm SM}$ is found using the same expression and $A_1 = 0$. The values of $k$, $k_{\rm SM}$, and $\theta_{\rm PC}$ for each simulation are collected in Table~1. 
%
 
\begin{table}
\centering 
\begin{tabular}{ l c c c c c}
  Metric               & $R_{\rm LC}/r_*$ & $a$        & $k_{\rm SM}$ & $k$ & $\theta_{\rm PC}$ (rads) \\
\hline
Kerr                  &    5          &  0.336    &  1.37  & 1.54 &  0.42 \\
Kerr                  &    10        &  0.168    &  1.34  & 1.50 & 0.29 \\
Schwarzschild  &    5          &  0           &  1.36  & 1.53 &  0.42 \\
Minkowski        &    5          &  ---         &   1.20  & 1.35 & 0.51 \\
Minkowski        &    10        &  ---         &   1.24  &  1.39 & 0.36 \\
\end{tabular}
\label{sim_table}
\caption{ Simulation parameters and basic properties of solutions }
\end{table}

As can be seen in Fig.~\ref{sim_fig}, the model accurately predicts the current structure near the polar axis, including the normalization of $J_\mu J^\mu/\rho c^2$ at the axis in Kerr spacetime. 
The angle $\theta_{\rm null}$, at which $J_\mu J^\mu = 0$, is recovered by the models with an error of $\delta\theta_{\rm null}/\theta_{\rm PC} =$ -0.037 (-0.028) for $R_{\rm LC}/r_* =$ 10 (5), where $\delta\theta_{\rm null} = \theta_{\rm null, model} - \theta_{\rm null, sim}$. The split-monopole model is more accurate in this regard, with errors of $\delta\theta_{\rm null}/\theta_{\rm PC} =$ 0.013 (0.027) for $R_{\rm LC}/r_* =$ 10 (5); here the slow-rotation model becomes more accurate as the stellar rotation rate is decreased. One the other hand, closer to the polar cap boundary $\theta_{\rm PC}$ the split-monopole model diverges somewhat from the simulations, as explained by the discussion in \S \ref{windzonegen}.
It is striking how closely both models follow the simulations even at these high spins---the two values investigated here correspond to spin frequencies of 477~Hz and 954~Hz, taking $r_* = 10$~km; the latter value corresponds to a pulsar faster than any yet discovered.

In the simulations, the return current has two components: a narrow, nearly singular current layer at the polar cap boundary, and a distributed return current which occupies a non-negligible volume. In Fig.~\ref{sim_fig} the distributed return current lies to the right of the curves' minima, which coincide with the colatitudes at which the contravariant vector component $J^r = 0$. In all simulations the width of the return-current region is approximately 11\% of the polar cap width, although this would increase if the field were non-dipolar and the polar cap were displaced from the polar axis. This is because the magnetic flux is typically more concentrated towards the outer edge of the polar cap for an on-axis polar cap as compared to an off-axis one in axisymmetry. The return-current region is slightly larger in the analytical models using the two-term poloidal flux approximation, extending over roughly 12\% of the polar cap width for the parameters shown.

The four-current structure in the Schwarzschild simulation is nearly indistinguishable from the flat spacetime results---clearly the frame dragging is the critical ingredient. One can understand this in the context of the model by noting that both $J_\mu J^\mu$ and $\rho^2$ are proportional to $1/\alpha^2$ [see equations~ (\ref{kerr_estimate}) and (\ref{kerr_gen})], and therefore the normalized four-current magnitude is independent of the lapse function.

\section{Discussion}
\label{discussion}
Assuming a force-free magnetic field structure, we have shown that there generically exist regions of timelike four-current with $0 < J_B/J_{\rm GJ}  < 1$ on the polar cap for an axisymmetric pulsar in both flat and Kerr spacetimes. Both theory \citep{Beloborodov_cloud} and one dimensional PIC simulations \citep{TimokhinArons} predict that pair production is not expected in these regions. 

Our analytical approach is valid for general magnetic field geometries, assuming the magnetic field varies on a scale larger than the scale of the polar cap. We show that in flat spacetime for a polar cap centered on the rotational axis, the four-current over the polar cap is timelike, except on the polar axis, where it is null. In Kerr spacetime, frame dragging reduces the Goldreich-Julian density over the polar cap and creates a spacelike four-current region near the polar axis. Additionally, if the polar cap is displaced from the rotational axis due to the presence of higher-order multipoles, there will generically be a region of spacelike four-current at its inner edge even in flat spacetime, unless the magnetic field is exactly vertical. 

We find it useful to consider the split-monopole model for the fields in the wind zone, since it allows one to derive simple and intuitive results. The split-monopole model is most accurate near the inner edge of the polar cap (the one closest to the polar axis). It predicts the highest latitude at which the four-current becomes null with a fractional error of $\lesssim 1\%$ for $R_{\rm LC} \gtrsim 10\, r_*$. 

However, force-free simulations show that there is a distributed return current region. For a surface dipole field, the distributed return current occupies $\sim 11 \%$ of the polar cap in latitude starting from the last open field line. Pair production occurs in this region, since the four current is timelike but is in the opposite sense as the Goldreich-Julian density flowing outward. 

Because the split-monopole model does not contain the distributed return current, we have considered a form of the poloidal field lines in the wind zone given by equation (\ref{poloidalB}). This form is obtained from a fit to force-free simulations by \citet{Tchekhovskoy_current}.  
Using their fitting formula, we find excellent agreement in the width of the return-current region between our analytical theory and force-free simulations. 


An interesting application of our results is to the hollow cone model of \citet{RadhakrishnanCooke}. In Kerr spacetime, pair-producing regions at the pole and in the distributed return current layer are separated by a timelike four-current region devoid of pair production. Thus, the spatial distribution of pair-producing regions resembles a hollow cone. If one associates radio emission with pair production, then our work provides a physical explanation for the hollow cone model, in those pulsars having nearly-aligned spin and magnetic axes.

In the case of significantly misaligned spin and magnetic axes, the axisymmetric approximation is severely violated. However, we point out that even in this case, it is still possible to trace the far-field distribution of current from the wind zone along magnetic field lines back to the polar cap. In particular, \citep{Gruzinov06} has shown that if the structure of magnetic field lines corotates with the star\footnote{The velocity of the field lines, which corotate rigidly, does not generally equal the drift velocity of the plasma.}, then under the force-free assumption in flat spacetime
\ba
\label{Gruzinoveq}
\bfnabla \times \left[\bfB +  \frac{\bfV_0}{c} \times  \left(\frac{\bfV_0}{c} \times \bfB\right)\right] = \lambda \bfB.
\ea
Here $\bfV_0$ is given by the flat spacetime expression in equation (\ref{V0eq}) and $\lambda$ is constant on magnetic field lines ($\bfB \cdot \bfnabla \lambda = 0$), which can be seen by taking the divergence of both sides of the equation. This means it is possible to generalize our methodology and trace back the current from the wind zone to determine the distribution of current over the polar cap in 3D.

Our results also have important implications for first-principles modeling of the pair cascade. They are broadly in agreement with the axisymmetric PIC simulations of \citet{PhilippovGR}. These authors find that in flat spacetime, pair production is shut off over the entire polar cap, whereas with general relativity there is still a large fraction of the polar cap at mid-latitudes that does not support pair production. Our results show that the features observed in these simulations are universal regardless of the magnetic field configuration or the ratio of the light cylinder radius to the neutron star radius. In fact, the only major assumption we require for our results to hold is that the magnetic field at the surface of the neutron star varies on scales larger than the size of the polar cap. 

Our results also suggest the presence of a ``gap" (a region where the density is below the Goldreich-Julian density) in the outer magnetosphere, due to the absence of surface pair production over a sizable fraction of the polar cap. However, we have only considered the $\gamma$--$B$ mechanism of pair production. A second channel for pair production that our analytical treatment does not consider is the photon-photon mechanism proposed by \citet{ChengHoRuderman}. The $\gamma$--$\gamma$ mechanism was simulated using PIC by \citet{ChenBeloborodov} and could potentially fill an outer gap with plasma. 

Nevertheless, even in the presence of $\gamma$--$\gamma$ pair production, one may still expect a thin gap above the return-current layer near the last open field line. This is due to the concave shape of the field lines in the closed zone of the magnetosphere, which makes it difficult for photons emitted in the region of open field lines or from the return current layer to fill the volume directly above the last open field line [see e.g.\ Fig. 3 of \cite{ChengHoRuderman}]. 

Gaps have been proposed as sites of pulsed high-energy emission \citep{Wattersetal,RomaniWatters,DyksRudak,MuslimovHarding}. However, alternatives that place the emission in the return current layer or current sheet also exist \citep{BaiSpitkovsky,Cerutti3D}. Thus, further computational work is needed to pin down theoretically the sites of pulsed high-energy emission.

\section*{Acknowledgements}
The authors would like to thank Jon Arons, Eliot Quataert, Anatoly Spitkovsky, Sasha Philippov, and Sam Gralla for stimulating discussions that helped to improve the paper. MB was supported by NASA Astrophysics Theory grant NNX14AH49G to the University of California, Berkeley and the Theoretical Astrophysics Center at UC Berkeley.  KP was supported by NASA through Einstein Postdoctoral Fellowship grant number PF5-160142 awarded by the Chandra X-ray Center, which is operated by the Smithsonian Astrophysical Observatory for NASA under contract NAS8-03060. This research used the SAVIO computational cluster resource provided by the Berkeley Research Computing program at the University of California, Berkeley (supported by the UC Berkeley Chancellor, Vice Chancellor of Research, and Office of the CIO).

\bibliography{../../bibliography/pulsar}

\begin{thebibliography}{66}
\expandafter\ifx\csname natexlab\endcsname\relax\def\natexlab#1{#1}\fi

\bibitem[{{Abdo} {et~al.}(2011){Abdo}, {Ackermann}, {Ajello}, {Allafort},
  {Baldini}, {Ballet}, {Barbiellini}, {Bastieri}, {Bechtol}, {Bellazzini},
  {Berenji}, {Blandford}, {Bloom}, {Bonamente}, {Borgland}, {Bouvier},
  {Brandt}, {Bregeon}, {Brez}, {Brigida}, {Bruel}, {Buehler}, {Buson},
  {Caliandro}, {Cameron}, {Cannon}, {Caraveo}, {Casandjian}, {{\c C}elik},
  {Charles}, {Chekhtman}, {Cheung}, {Chiang}, {Ciprini}, {Claus},
  {Cohen-Tanugi}, {Costamante}, {Cutini}, {D'Ammando}, {Dermer}, {de Angelis},
  {de Luca}, {de Palma}, {Digel}, {do Couto e Silva}, {Drell}, {Drlica-Wagner},
  {Dubois}, {Dumora}, {Favuzzi}, {Fegan}, {Ferrara}, {Focke}, {Fortin},
  {Frailis}, {Fukazawa}, {Funk}, {Fusco}, {Gargano}, {Gasparrini}, {Gehrels},
  {Germani}, {Giglietto}, {Giordano}, {Giroletti}, {Glanzman}, {Godfrey},
  {Grenier}, {Grondin}, {Grove}, {Guiriec}, {Hadasch}, {Hanabata}, {Harding},
  {Hayashi}, {Hayashida}, {Hays}, {Horan}, {Itoh}, {J{\'o}hannesson},
  {Johnson}, {Johnson}, {Khangulyan}, {Kamae}, {Katagiri}, {Kataoka}, {Kerr},
  {Kn{\"o}dlseder}, {Kuss}, {Lande}, {Latronico}, {Lee}, {Lemoine-Goumard},
  {Longo}, {Loparco}, {Lubrano}, {Madejski}, {Makeev}, {Marelli}, {Mazziotta},
  {McEnery}, {Michelson}, {Mitthumsiri}, {Mizuno}, {Moiseev}, {Monte},
  {Monzani}, {Morselli}, {Moskalenko}, {Murgia}, {Nakamori}, {Naumann-Godo},
  {Nolan}, {Norris}, {Nuss}, {Ohsugi}, {Okumura}, {Omodei}, {Ormes}, {Ozaki},
  {Paneque}, {Parent}, {Pelassa}, {Pepe}, {Pesce-Rollins}, {Pierbattista},
  {Piron}, {Porter}, {Rain{\`o}}, {Rando}, {Ray}, {Razzano}, {Reimer},
  {Reimer}, {Reposeur}, {Ritz}, {Romani}, {Sadrozinski}, {Sanchez},
  {Parkinson}, {Scargle}, {Schalk}, {Sgr{\`o}}, {Siskind}, {Smith}, {Spandre},
  {Spinelli}, {Strickman}, {Suson}, {Takahashi}, {Takahashi}, {Tanaka},
  {Thayer}, {Thompson}, {Tibaldo}, {Torres}, {Tosti}, {Tramacere}, {Troja},
  {Uchiyama}, {Vandenbroucke}, {Vasileiou}, {Vianello}, {Vitale}, {Wang},
  {Wood}, {Yang}, \& {Ziegler}}]{Fermi2}
{Abdo}, A.~A., {Ackermann}, M., {Ajello}, M., {et~al.} 2011, Science, 331, 739

\bibitem[{{Abdo} {et~al.}(2013){Abdo}, {Ajello}, {Allafort}, {Baldini},
  {Ballet}, {Barbiellini}, {Baring}, {Bastieri}, {Belfiore}, {Bellazzini}, \&
  et~al.}]{Fermia}
{Abdo}, A.~A., {Ajello}, M., {Allafort}, A., {et~al.} 2013, \apjs, 208, 17

\bibitem[{{Aliu} {et~al.}(2008){Aliu}, {Anderhub}, {Antonelli}, {Antoranz},
  {Backes}, {Baixeras}, {Barrio}, {Bartko}, {Bastieri}, {Becker}, {Bednarek},
  {Berger}, {Bernardini}, {Bigongiari}, {Biland}, {Bock}, {Bonnoli}, {Bordas},
  {Bosch-Ramon}, {Bretz}, {Britvitch}, {Camara}, {Carmona}, {Chilingarian},
  {Commichau}, {Contreras}, {Cortina}, {Costado}, {Covino}, {Curtef}, {Dazzi},
  {De Angelis}, {De Cea del Pozo}, {de los Reyes}, {De Lotto}, {De Maria}, {De
  Sabata}, {Delgado Mendez}, {Dominguez}, {Dorner}, {Doro}, {Els{\"a}sser},
  {Errando}, {Fagiolini}, {Ferenc}, {Fernandez}, {Firpo}, {Fonseca}, {Font},
  {Galante}, {Garcia Lopez}, {Garczarczyk}, {Gaug}, {Goebel}, {Hadasch},
  {Hayashida}, {Herrero}, {H{\"o}hne}, {Hose}, {Hsu}, {Huber}, {Jogler},
  {Kranich}, {La Barbera}, {Laille}, {Leonardo}, {Lindfors}, {Lombardi},
  {Longo}, {Lopez}, {Lorenz}, {Majumdar}, {Maneva}, {Mankuzhiyil}, {Mannheim},
  {Maraschi}, {Mariotti}, {Martinez}, {Mazin}, {Meucci}, {Meyer}, {Miranda},
  {Mirzoyan}, {Moles}, {Moralejo}, {Nieto}, {Nilsson}, {Ninkovic}, {Otte},
  {Oya}, {Paoletti}, {Paredes}, {Pasanen}, {Pascoli}, {Pauss}, {Pegna},
  {Perez-Torres}, {Persic}, {Peruzzo}, {Piccioli}, {Prada}, {Prandini},
  {Puchades}, {Raymers}, {Rhode}, {Rib{\'o}}, {Rico}, {Rissi}, {Robert},
  {R{\"u}gamer}, {Saggion}, {Saito}, {Salvati}, {Sanchez-Conde}, {Sartori},
  {Satalecka}, {Scalzotto}, {Scapin}, {Schweizer}, {Shayduk}, {Shinozaki},
  {Shore}, {Sidro}, {Sierpowska-Bartosik}, {Sillanp{\"a}{\"a}}, {Sobczynska},
  {Spanier}, {Stamerra}, {Stark}, {Takalo}, {Tavecchio}, {Temnikov}, {Tescaro},
  {Teshima}, {Tluczykont}, {Torres}, {Turini}, {Vankov}, {Venturini}, {Vitale},
  {Wagner}, {Wittek}, {Zabalza}, {Zandanel}, {Zanin}, {Zapatero}, {de Jager},
  {de Ona Wilhelmi}, \& {MAGIC Collaboration}}]{MAGICa}
{Aliu}, E., {Anderhub}, H., {Antonelli}, L.~A., {et~al.} 2008, Science, 322,
  1221

\bibitem[{{Arons}(1979)}]{Arons79}
{Arons}, J. 1979, \ssr, 24, 437

\bibitem[{{Arons}(2012)}]{Arons_review}
---. 2012, \ssr, 173, 341

\bibitem[{{Arons} \& {Scharlemann}(1979)}]{AronsScharlemann}
{Arons}, J., \& {Scharlemann}, E.~T. 1979, \apj, 231, 854

\bibitem[{{Bai} \& {Spitkovsky}(2010)}]{BaiSpitkovsky}
{Bai}, X.-N., \& {Spitkovsky}, A. 2010, \apj, 715, 1282

\bibitem[{{Beloborodov}(2008)}]{Beloborodov_cloud}
{Beloborodov}, A.~M. 2008, \apjl, 683, L41

\bibitem[{{Belyaev}(2015{\natexlab{a}})}]{BelyaevPIC2}
{Belyaev}, M.~A. 2015{\natexlab{a}}, \mnras, 449, 2759

\bibitem[{{Belyaev}(2015{\natexlab{b}})}]{BelyaevPIC}
---. 2015{\natexlab{b}}, New Astronomy, 36, 37

\bibitem[{{Beskin}(1990)}]{BeskinGR}
{Beskin}, V.~S. 1990, Soviet Astronomy Letters, 16, 286

\bibitem[{{Blandford} \& {Znajek}(1977)}]{BlandfordZnajek}
{Blandford}, R.~D., \& {Znajek}, R.~L. 1977, \mnras, 179, 433

\bibitem[{{Bucciantini} {et~al.}(2011){Bucciantini}, {Arons}, \&
  {Amato}}]{Bucciantini_nebula}
{Bucciantini}, N., {Arons}, J., \& {Amato}, E. 2011, \mnras, 410, 381

\bibitem[{{Cerutti} {et~al.}(2014){Cerutti}, {Philippov}, {Parfrey}, \&
  {Spitkovsky}}]{CeruttiSpitkovsky}
{Cerutti}, B., {Philippov}, A., {Parfrey}, K., \& {Spitkovsky}, A. 2014,
  arXiv:1410.3757

\bibitem[{{Cerutti} {et~al.}(2015){Cerutti}, {Philippov}, \&
  {Spitkovsky}}]{Cerutti3D}
{Cerutti}, B., {Philippov}, A.~A., \& {Spitkovsky}, A. 2015, ArXiv e-prints

\bibitem[{{Chen} \& {Beloborodov}(2014)}]{ChenBeloborodov}
{Chen}, A.~Y., \& {Beloborodov}, A.~M. 2014, \apjl, 795, L22

\bibitem[{{Cheng} {et~al.}(1986){Cheng}, {Ho}, \& {Ruderman}}]{ChengHoRuderman}
{Cheng}, K.~S., {Ho}, C., \& {Ruderman}, M. 1986, \apj, 300, 500

\bibitem[{{Contopoulos} \& {Kalapotharakos}(2010)}]{CK}
{Contopoulos}, I., \& {Kalapotharakos}, C. 2010, \mnras, 404, 767

\bibitem[{{Contopoulos} {et~al.}(1999){Contopoulos}, {Kazanas}, \&
  {Fendt}}]{Contopoulos}
{Contopoulos}, I., {Kazanas}, D., \& {Fendt}, C. 1999, \apj, 511, 351

\bibitem[{{Daugherty} \& {Harding}(1982)}]{DaughertyHarding}
{Daugherty}, J.~K., \& {Harding}, A.~K. 1982, \apj, 252, 337

\bibitem[{{de Jager} {et~al.}(1996){de Jager}, {Harding}, {Michelson}, {Nel},
  {Nolan}, {Sreekumar}, \& {Thompson}}]{deJager}
{de Jager}, O.~C., {Harding}, A.~K., {Michelson}, P.~F., {et~al.} 1996, \apj,
  457, 253

\bibitem[{{Dyks} \& {Rudak}(2003)}]{DyksRudak}
{Dyks}, J., \& {Rudak}, B. 2003, \apj, 598, 1201

\bibitem[{{Goldreich} \& {Julian}(1969)}]{GoldreichJulian}
{Goldreich}, P., \& {Julian}, W.~H. 1969, \apj, 157, 869

\bibitem[{{Gralla} {et~al.}(2016){Gralla}, {Lupsasca}, \& {Philippov}}]{Gralla}
{Gralla}, S.~E., {Lupsasca}, A., \& {Philippov}, A. 2016, ArXiv e-prints

\bibitem[{{Gruzinov}(2005)}]{Gruzinov}
{Gruzinov}, A. 2005, Physical Review Letters, 94, 021101

\bibitem[{{Gruzinov}(2006)}]{Gruzinov06}
---. 2006, \apjl, 647, L119

\bibitem[{{Harding} \& {Muslimov}(1998)}]{HardingMuslimov1998}
{Harding}, A.~K., \& {Muslimov}, A.~G. 1998, \apj, 508, 328

\bibitem[{{Harding} \& {Muslimov}(2011)}]{HardingMuslimov11}
---. 2011, \apj, 743, 181

\bibitem[{{Hibschman} \& {Arons}(2001)}]{HibschmanArons}
{Hibschman}, J.~A., \& {Arons}, J. 2001, \apj, 560, 871

\bibitem[{{Kalapotharakos} {et~al.}(2014){Kalapotharakos}, {Harding}, \&
  {Kazanas}}]{Kalapotharakos_FIDO}
{Kalapotharakos}, C., {Harding}, A.~K., \& {Kazanas}, D. 2014, \apj, 793, 97

\bibitem[{{Kalapotharakos} {et~al.}(2012){Kalapotharakos}, {Kazanas},
  {Harding}, \& {Contopoulos}}]{Kalapotharakos}
{Kalapotharakos}, C., {Kazanas}, D., {Harding}, A., \& {Contopoulos}, I. 2012,
  \apj, 749, 2

\bibitem[{{Komissarov}(2002)}]{KomissarovFFE}
{Komissarov}, S.~S. 2002, \mnras, 336, 759

\bibitem[{{Komissarov}(2006)}]{Komissarov}
---. 2006, \mnras, 367, 19

\bibitem[{{Komissarov}(2011)}]{KomissarovGR}
---. 2011, \mnras, 418, L94

\bibitem[{{Li} {et~al.}(2012){Li}, {Spitkovsky}, \&
  {Tchekhovskoy}}]{LiSpitkovsky}
{Li}, J., {Spitkovsky}, A., \& {Tchekhovskoy}, A. 2012, \apj, 746, 60

\bibitem[{{Lyutikov}(2011)}]{Lyutikov}
{Lyutikov}, M. 2011, \prd, 83, 124035

\bibitem[{{MacDonald} \& {Thorne}(1982)}]{MT82}
{MacDonald}, D., \& {Thorne}, K.~S. 1982, \mnras, 198, 345

\bibitem[{{McKinney}(2006)}]{McKinney}
{McKinney}, J.~C. 2006, \mnras, 368, L30

\bibitem[{{Mestel} {et~al.}(1985){Mestel}, {Robertson}, {Wang}, \&
  {Westfold}}]{Mestel_cloud}
{Mestel}, L., {Robertson}, J.~A., {Wang}, Y.-M., \& {Westfold}, K.~C. 1985,
  \mnras, 217, 443

\bibitem[{{Michel}(1973)}]{Michelmono}
{Michel}, F.~C. 1973, \apjl, 180, L133

\bibitem[{{Muslimov} \& {Harding}(2004)}]{MuslimovHarding}
{Muslimov}, A.~G., \& {Harding}, A.~K. 2004, \apj, 606, 1143

\bibitem[{{Muslimov} \& {Tsygan}(1992)}]{MuslimovTsygan}
{Muslimov}, A.~G., \& {Tsygan}, A.~I. 1992, \mnras, 255, 61

\bibitem[{{Parfrey} {et~al.}(2012){Parfrey}, {Beloborodov}, \&
  {Hui}}]{Parfreyetal}
{Parfrey}, K., {Beloborodov}, A.~M., \& {Hui}, L. 2012, \mnras, 423, 1416

\bibitem[{{Parfrey}(2012)}]{ParfreyPhDT}
{Parfrey}, K.~P. 2012, PhD thesis, Columbia University

\bibitem[{{P{\'e}tri}(2012)}]{PetriFF}
{P{\'e}tri}, J. 2012, \mnras, 424, 605

\bibitem[{{P{\'e}tri}(2016)}]{PetriGR}
---. 2016, \mnras, 455, 3779

\bibitem[{{Philippov} {et~al.}(2015){Philippov}, {Cerutti}, {Tchekhovskoy}, \&
  {Spitkovsky}}]{PhilippovGR}
{Philippov}, A.~A., {Cerutti}, B., {Tchekhovskoy}, A., \& {Spitkovsky}, A.
  2015, ArXiv e-prints

\bibitem[{{Philippov} \& {Spitkovsky}(2014)}]{PhilippovSpitkovsky}
{Philippov}, A.~A., \& {Spitkovsky}, A. 2014, \apjl, 785, L33

\bibitem[{{Philippov} {et~al.}(2014){Philippov}, {Spitkovsky}, \&
  {Cerutti}}]{PhilippovSpitkovsky1}
{Philippov}, A.~A., {Spitkovsky}, A., \& {Cerutti}, B. 2014, arXiv:1412.0673

\bibitem[{{Radhakrishnan} \& {Cooke}(1969)}]{RadhakrishnanCooke}
{Radhakrishnan}, V., \& {Cooke}, D.~J. 1969, \aplett, 3, 225

\bibitem[{{Rankin}(1990)}]{Rankin}
{Rankin}, J.~M. 1990, \apj, 352, 247

\bibitem[{{Ravenhall} \& {Pethick}(1994)}]{RavenhallPethick}
{Ravenhall}, D.~G., \& {Pethick}, C.~J. 1994, \apj, 424, 846

\bibitem[{{Romani} \& {Watters}(2010)}]{RomaniWatters}
{Romani}, R.~W., \& {Watters}, K.~P. 2010, \apj, 714, 810

\bibitem[{{Shklovsky}(1968)}]{Shklovsky}
{Shklovsky}, J.~S. 1968, {Supernovae}

\bibitem[{{Spitkovsky}(2006)}]{Spitkovsky}
{Spitkovsky}, A. 2006, \apjl, 648, L51

\bibitem[{{Sturrock}(1971)}]{Sturrock}
{Sturrock}, P.~A. 1971, \apj, 164, 529

\bibitem[{{Tavani} {et~al.}(2011){Tavani}, {Bulgarelli}, {Vittorini},
  {Pellizzoni}, {Striani}, {Caraveo}, {Weisskopf}, {Tennant}, {Pucella},
  {Trois}, {Costa}, {Evangelista}, {Pittori}, {Verrecchia}, {Del Monte},
  {Campana}, {Pilia}, {De Luca}, {Donnarumma}, {Horns}, {Ferrigno}, {Heinke},
  {Trifoglio}, {Gianotti}, {Vercellone}, {Argan}, {Barbiellini}, {Cattaneo},
  {Chen}, {Contessi}, {D'Ammando}, {DeParis}, {Di Cocco}, {Di Persio},
  {Feroci}, {Ferrari}, {Galli}, {Giuliani}, {Giusti}, {Labanti}, {Lapshov},
  {Lazzarotto}, {Lipari}, {Longo}, {Fuschino}, {Marisaldi}, {Mereghetti},
  {Morelli}, {Moretti}, {Morselli}, {Pacciani}, {Perotti}, {Piano}, {Picozza},
  {Prest}, {Rapisarda}, {Rappoldi}, {Rubini}, {Sabatini}, {Soffitta},
  {Vallazza}, {Zambra}, {Zanello}, {Lucarelli}, {Santolamazza}, {Giommi},
  {Salotti}, \& {Bignami}}]{AGILE}
{Tavani}, M., {Bulgarelli}, A., {Vittorini}, V., {et~al.} 2011, Science, 331,
  736

\bibitem[{{Tchekhovskoy} {et~al.}(2016){Tchekhovskoy}, {Philippov}, \&
  {Spitkovsky}}]{Tchekhovskoy_current}
{Tchekhovskoy}, A., {Philippov}, A., \& {Spitkovsky}, A. 2016, \mnras, 457,
  3384

\bibitem[{{Tchekhovskoy} {et~al.}(2013){Tchekhovskoy}, {Spitkovsky}, \&
  {Li}}]{Tchekhovskoy}
{Tchekhovskoy}, A., {Spitkovsky}, A., \& {Li}, J.~G. 2013, \mnras, 435, L1

\bibitem[{{Thorne} \& {MacDonald}(1982)}]{TM82}
{Thorne}, K.~S., \& {MacDonald}, D. 1982, \mnras, 198, 339

\bibitem[{{Thorne} {et~al.}(1986){Thorne}, {Price}, \& {MacDonald}}]{membrane}
{Thorne}, K.~S., {Price}, R.~H., \& {MacDonald}, D.~A. 1986, {Black holes: The
  membrane paradigm}

\bibitem[{{Timokhin}(2006)}]{Timokhin_forcefree}
{Timokhin}, A.~N. 2006, \mnras, 368, 1055

\bibitem[{{Timokhin} \& {Arons}(2013)}]{TimokhinArons}
{Timokhin}, A.~N., \& {Arons}, J. 2013, \mnras, 429, 20

\bibitem[{{Timokhin} \& {Harding}(2015)}]{Timokhin_max}
{Timokhin}, A.~N., \& {Harding}, A.~K. 2015, \apj, 810, 144

\bibitem[{{Wasserman} \& {Shapiro}(1983)}]{WassermanShapiro}
{Wasserman}, I., \& {Shapiro}, S.~L. 1983, \apj, 265, 1036

\bibitem[{{Watters} {et~al.}(2009){Watters}, {Romani}, {Weltevrede}, \&
  {Johnston}}]{Wattersetal}
{Watters}, K.~P., {Romani}, R.~W., {Weltevrede}, P., \& {Johnston}, S. 2009,
  \apj, 695, 1289

\end{thebibliography}

\end{document}